\documentclass[RNAAS]{aastex631}
\usepackage{amsmath}
\usepackage{graphicx}

\begin{document}
	
\title{Theoretical calculation of the  antenna impedance and  shot noise \\
	at low-frequencies: application to Parker Solar Probe}

\correspondingauthor{Nicole Meyer-Vernet}
\email{nicole.meyer@obspm.fr}

\author[0000-0001-6449-5274]{Nicole Meyer-Vernet}
\affiliation{LIRA, Observatoire de Paris, PSL Université,\\ CNRS, Sorbonne Université, Université Paris Cité, 92195 Meudon, France}
\author[0009-0000-6187-4463]{Baptiste Verkampt}
\affiliation{LIRA, Observatoire de Paris, PSL Université,\\ CNRS, Sorbonne Université, Université Paris Cité, 92195 Meudon, France}
\author[0000-0002-0287-9229]{Pietro 	Dazzi}
\affiliation{Centre for Mathematical Plasma Astrophysics, Department of Mathematics, KU Leuven, Celestijnenlaan 200B, 3001 Leuven, Belgium}
\affiliation{LIRA, Observatoire de Paris, PSL Université,\\ CNRS, Sorbonne Université, Université Paris Cité, 92195 Meudon, France}
\author[0000-0002-2757-101X]{Karine Issautier}
\affiliation{LIRA, Observatoire de Paris, PSL Université,\\ CNRS, Sorbonne Université, Université Paris Cité, 92195 Meudon, France}

\begin{abstract}
	 The voltage power spectral density measured   around the ambient plasma frequency in space is not affected by spacecraft perturbations that impact  traditional plasma analysers. The spectroscopy of this  noise, produced by the quasi-thermal motion of  ambient charged particles, is thus an efficient tool for measuring  in situ plasma properties in space.  In contrast, the spectrum at lower frequencies, which is determined by  the parallel antenna resistance  due to  electric currents,  depends on the spacecraft local environment.  Recently, \citet{zhe26}  erroneously estimated this  resistance from Parker Solar Probe (PSP) data. We hereby  present a theoretical calculation of this resistance, which determines  the shot noise and the receiver gain at low frequencies, and provide a preliminary comparison to  PSP/FIELDS data. 	  We also show that this resistance can change  the receiver gain in the frequency range used for QTN spectroscopy during PSP  inner orbits.
\end{abstract}


\section{Introduction}
Quasi-thermal noise (QTN) spectroscopy is routinely used to measure in situ  plasma  properties in space with sensitive voltage receivers   (\citep{mey86, mey98, mey17, mon20} and references therein). In weakly magnetised plasmas, the electron QTN consists of a plateau below the plasma frequency $f_p$,  produced by electrons crossing the plasma sheath surrounding the antenna, a peak close to $f_p$,  produced by electrons of speed close to the large Langmuir wave phase speed, and a power  decreasing as large frequencies \citep{mey89}. 

Since the spectrum around $f_p$ is based on waves of large wavelength, this technique is  equivalent to a detector of  cross section much larger than that  of  particle  analysers, and is  relatively immune to spacecraft  charging effects which affect them. 
In contrast, the low-frequency spectrum ($f \ll f_p$), which is mainly determined by the shot noise produced by the currents through the antenna, is strongly affected by the spacecraft close environment.

This shot noise \citep{pet75, mey83, mey89} has been previously studied at frequencies   used for QTN studies  \citep{mey17}. However, on Parker Solar Probe close to the Sun, the large ambient electron density makes the plasma frequency very high, so the shot noise can be measured at frequencies   $f \ll f_p$ at which the parallel antenna resistance due to the currents plays an important role. This noise was recently studied by \citet{zhe26}. However, this paper does not calculate correctly the low-frequency antenna resistance.

\section{Antenna parallel resistance at low frequencies}

Consider a dipole antenna made of two arms of length $L$ and radius $a$. All quantities are in S.I. units, except the temperatures, in eV. At frequencies just below $f_p$, the  impedance mainly consists of a capacitance $C \simeq \pi \epsilon_0 L/\ln (L_D/a)$   for $L\gg L_D$, the Debye length \citep{mey89}, since the resistance  in series $V_{\mathrm{ThermalNoise}}^2/4eT_e$ (from Nyquist theorem) is generally negligible. However, for   $f \ll f_p$, the resistive component in parallel due to the  currents  plays an important role. It is  approximately given by $R = 1/\left|dI/d/\Phi  \right|$, where $ dI/d\Phi $ is the derivative of the current with respect to the antenna potential (see for example \citep{hen11}). In the absence of biasing, the current mainly consists of ejected photoelectrons of temperature $T_{ph}$, $I_{ph}  \propto \exp (-\Phi /T_{ph})$  and collected ambient electrons  of temperature $T$, $I_e$, since the ion current is negligible because $m_i\gg m_e$. Since the photoelectron current is the fastest charging process \citep{mey07}, we have  $1/R \simeq \left| dI_{ph}/d\Phi  \right|$, which reduces to $1/R \simeq  \left| I_e/ T_{ph} \right| $ since  $dI_{ph}/d\Phi  = - I_{ph}/ T_{ph}$ and  $I_e \simeq    I_{ph} $ at  current equilibrium in the absence of biasing currents.  The ambient electron current to  both antenna arms $I_e \simeq 2 e N_e$   where
\begin{equation}
	N_e \simeq  n S (eT/(2\pi m))^{1/2} \alpha  \label{Ne}
\end{equation}
is the plasma electron flux  on one arm, $S =  2 \pi a L$, $n$ and $T$ the electron density and temperature, $e$ and $m$ the electron charge and mass and $\alpha$ the correction due to the positive antenna potential \citep{laf73}. The antenna low-frequency resistance in parallel is thus
\begin{equation}
R \simeq T_{ph}/(2  e N_e)  \label{R}
\end{equation}
The low-frequency antenna impedance is  $Z = R/(1-iRC \omega)$
and  the shot noise is given by \citep{mey89}
\begin{equation}
V_{\mathrm{shot}}^2  \simeq 2 e^2 N_e \left| Z \right|^2 \Gamma ^2 
\end{equation}

Since the receiver impedance is mainly the base capacitance, i.e. $Z_R \simeq i/C_b \omega$, we deduce the receiver gain $\Gamma ^2$ and the shot noise for $f<f_p$

\begin{equation}
	\Gamma ^2  = \left| Z_R/(Z_R +Z) \right| ^2 \simeq \frac{1+R^2 C^2 \omega ^2}{1+R^2 (C+C_b)^2 \omega ^2} \label{gamma2}
\end{equation}
\begin{equation}
	V_{\mathrm{shot}}^2  \simeq   2 e^2 N_e \left[  \frac{R^2}{1+R^2 (C+C_b)^2 \omega ^2} \right] \label{VI}
\end{equation}
Equations (\ref{gamma2}) and (\ref{VI}) show that the  resistance $R$ increases the gain and flattens the spectrum, which is no longer proportional to $1/\omega ^2$ for  $f \leq 1/(2 \pi R (C+C_b))$. Note that most publications drawing 
the theoretical shot noise  apply for  $f>f_p$ the expression calculated by \citep{mey89} for  $f<f_p$. This is incorrect since the  variation $\propto 1/f^2$ stems from the squared  Fourier transform of a  Heaviside function, implying that the  signal produced by a collected electron varies quicker than $1/f$, which is erroneous for $f>f_p$ \citep{mey83}.
 
\begin{figure*}
	\centering
	\includegraphics[width=12cm]{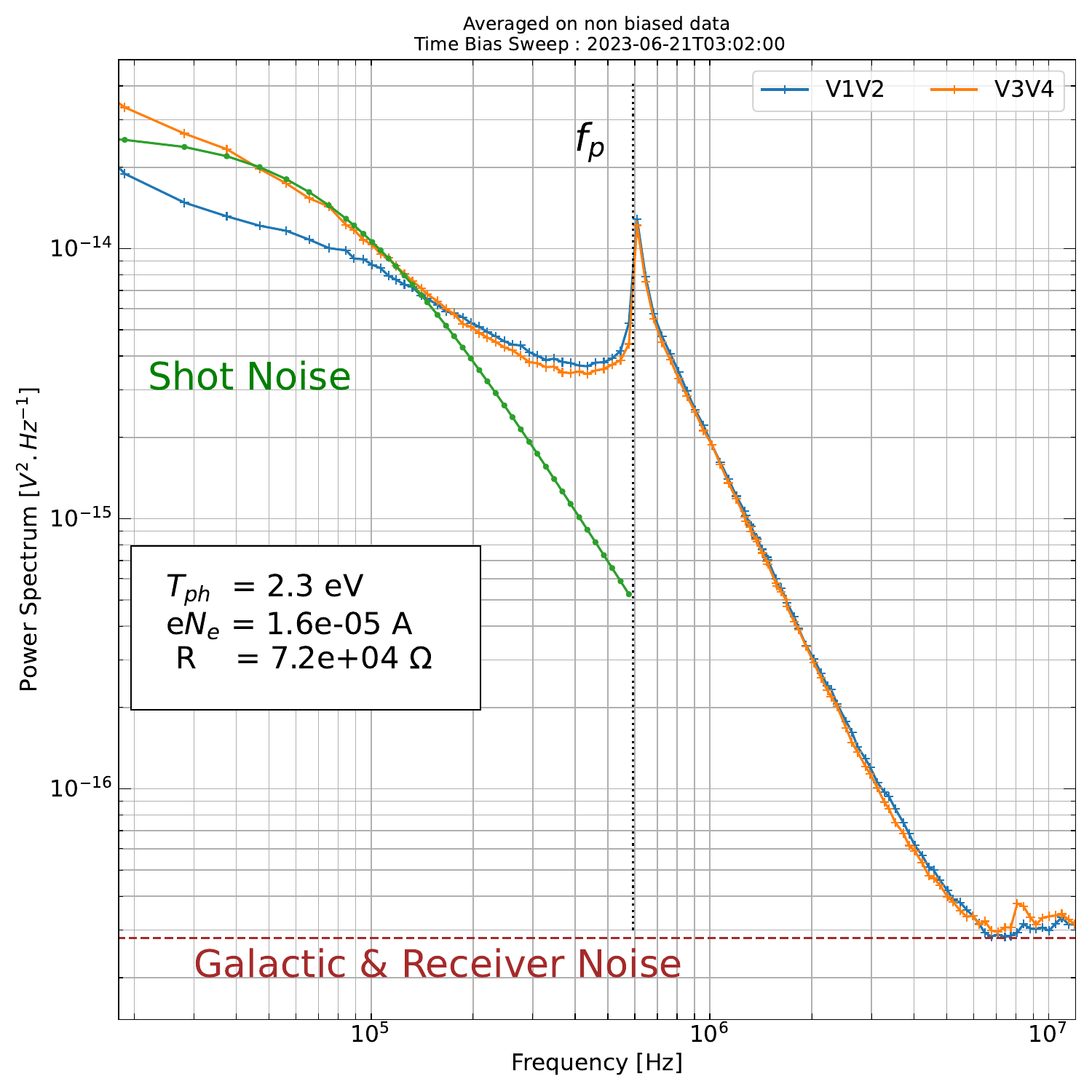}
	\caption{Theoretical shot noise (in green) superimposed to the power spectrum  measured in the absence of biasing on  the two PSP/FIELDS dipoles (in respectively orange and blue) 
		at heliocentric distance $\simeq 19 R_s$. The theoretical shot noise, which largely  dominates the electron QTN  below 100 kHz, roughly agrees with the  low-frequency spectrum measured by the unperturbed dipole V3V4.}
	\label{Fig1}
\end{figure*}

\section{Application to PSP}

Let us  apply these results to data from PSP/FIELDS \citep{bal16}. Figure 1 shows the spectral power density measured by the  dipoles V1V2 and V3V4 in the absence of biasing at  heliocentric distance $\simeq 19 R_s$. The spectra for both dipoles are similar for $f> 20 $ kHz, but differ significantly  at lower frequencies. There is a systematic decrease in the low-frequency power on the V1V2 dipole, that began in  orbit 8 due to a perturbation in  the circuit of that antenna.

From the QTN spectral density, we deduce $n \simeq 4.4 \times 10^9 $ m$^{-3}$, $T\simeq 40 $ eV, thus $L_D\simeq 0.9 $ m, $C\simeq 9 \times 10^{-12}$ F. Therefore, with $\alpha \simeq 1.1$ for PSP/FIELDS antennas at this heliocentric distance, Eq.(\ref{Ne}) yields $eN_e \simeq 1.6 \times  10^{-5}$ A. The  resistance $R$ given by (\ref{R}) is proportional to $T_{ph}/(nT^{1/2})$. We estimate the photoelectron temperature $T_{ph} \simeq 2.3 $ eV, not far from the values used by \citet{erg10} and \citet{mar14}.  We deduce from  (\ref{R}) $R \simeq 7.2  \times 10^4$ ohms. The corresponding shot noise  (\ref{VI}) with   $C_b\simeq 18 \times  10^{-12}$ F is plotted  on Figure 1.

\section{Conclusion}

We have calculated the parallel resistance due to the currents through the antenna in the absence of bias. The  agreement with the  low-frequency data (for which  the shot noise dominates the QTN) should be confirmed by  accurate estimates of the total noise, including  the ion contribution  \citep{iss96} with more data. Equation (\ref{gamma2}) shows that the resistance changes considerably the receiver gain when $RC \omega \leq 1$. Since $R\propto 1/(nT^{1/2})$, with $n T^{1/2} \propto r^{-2.3}$ and C varies weakly with distance, this occurs when $f/f_p \leq  0.3 \times  (19/r_{\mathrm{R_s}})^{1.3} $, and therefore might affect QTN diagnostics close to the Sun. Our calculations, which can be easily generalised to  biased antennas, can be improved by including the receiver conductance and additional datasets in the absence of biasing currents, accurate plasma properties from QTN spectroscopy and more accurate estimates of the photoelectron temperature (Dazzi et al. 2026, in preparation).
\vspace{1cm}


\end{document}